\documentclass[12pt,preprint]{emulateapj}
\submitted{Submitted 2009 July 14; accepted 2009 September 14}
\journalinfo{The Astrophysical Journal, in press}

\newif\ifjournal\journalfalse

\usepackage{graphicx}
\usepackage{bm}
\renewcommand{\vec}[1]{\boldsymbol{#1}}
\newcommand{\grad}{\nabla}
\renewcommand{\div}{\nabla \cdot}
\newcommand{\rot}{\nabla \times}
\newcommand{\pp}[2]{\frac{\partial #1}{\partial #2}}
\def\pop{Phys. Plasmas}
\def\jcp{J.~Comput.~Phys.}

\shorttitle{RELATIVISTIC TWO-FLUID SIMULATIONS OF GUIDE FIELD RECONNECTION}
\shortauthors{ZENITANI, HESSE, \& KLIMAS}

\begin{document}

\title{Relativistic Two-fluid Simulations of Guide Field Magnetic Reconnection}

\author{Seiji Zenitani, Michael Hesse, and Alex Klimas}
\affil{
NASA Goddard Space Flight Center, Greenbelt, MD 20771, USA;
Seiji.Zenitani-1@nasa.gov
}

\begin{abstract}
The nonlinear evolution of relativistic magnetic reconnection
in sheared magnetic configuration (with a guide field)
is investigated by using
two-dimensional relativistic two-fluid simulations.
Relativistic guide field reconnection features
the charge separation and the guide field compression
in and around the outflow channel.
As the guide field increases,
the composition of the outgoing energy changes
from enthalpy-dominated to Poynting-dominated.
The inertial effects of the two-fluid model
play an important role to sustain magnetic reconnection.
Implications for the single-fluid magnetohydrodynamic approach and
the physics models of relativistic reconnection are briefly addressed.
\end{abstract}


\section{INTRODUCTION}

The role of magnetic fields has been recognized
in various contexts of relativistic astrophysics:
pulsars, magnetars, gamma-ray bursts (GRBs),
active galactic nuclei (AGNs), and black holes.
Owning to its fundamental nature,
relativistic magnetic reconnection
in electron--positron pair plasmas (or in pairs and baryons)
has drawn attentions in these sites as well;
however, its mechanisms are poorly understood.
Despite several theoretical attempts over decades \citep{bf94b,lyut03b,lyu05},
the most important properties of relativistic magnetic reconnection
such as the energy conversion rate and the released energy composition
are still under debate,
especially in the magnetically dominated limit.

Relativistic magnetohydrodynamic (RMHD) codes are very desirable to
study the theories of relativistic magnetic reconnection
and to investigate astrophysical problems which involve magnetic reconnection.
However, surprisingly,
reconnection or non-ideal RMHD problems were left untouched
until \citet{naoyuki06} carried out resistive RMHD simulations.
They exhibited a Petschek-like structure
with an Alfv\'{e}nic outflow
and showed that relativistic reconnection is
faster than the nonrelativistic counterpart.
Several groups have developed
resistive RMHD codes \citep{kom07,pal09,dumbser09},
which can be applied to the reconnection problems as well. 
Recently, \citet{zeni09a} employed
a relativistic two-fluid model to simulate magnetic reconnection.
They obtained Petschek-type steady reconnection in a large system
and found that the reconnection speed becomes
faster and faster in magnetically dominated regimes. 

In addition to the basic antiparallel configuration,
reconnection with a current-aligned magnetic field (the ``guide field'')
is likely in magnetic shear and in celestial flare situations.
It is well known that
guide field reconnection behaves differently
from its anti-parallel counterpart.
In the relativistic regime,
it is expected to change the energy composition
in the outflow in RMHD scales \citep{lyu05}.
However, the RMHD/fluid-scale behavior of
relativistic guide field reconnection
has not yet been explored by simulations.

In kinetic scales, it has been found that
reconnection is a powerful particle accelerator
in both antiparallel and guide field configurations
\citep{zeni01,zeni07,zeni08,claus04}.
Since the plasma nonthernal energy
is comparable with or exceeds the thermal part,
we expect that kinetic physics significantly affects the global dynamics.
Comparison with RMHD/fluid models
is useful to understand the role of kinetic physics.

The purpose of this paper is to advance
our two-fluid reconnection work \citep{zeni09a}
another step forward.
In the earlier work,
we assumed symmetric motions of the two species,
which enforced the charge neutrality in the system.
In this work, we solve two fluid motions independently.
This allows us to explore broader ranges of physical targets,
which involve charge separation.
Using relativistic full two-fluid simulations,
we will investigate a nonlinear development of
a relativistic reconnection system with a guide field.

\section{NUMERICAL SETUP}

We use the following set of relativistic fluid equations and Maxwell equations.
The subscript $s$ denotes species (`$p$' for positrons and `$e$' for electrons).
Similar equations for electrons are considered too:
\begin{equation}
\pp{}{t} \gamma_p n_p = -\div (n_p \vec{u}_p), \\
\end{equation}
\begin{eqnarray}
\label{eq:mom}
&& \pp{}{t} \Big( \frac{ \gamma_p w_p \vec{u}_p }{c^2} \Big)
= -\div \Big( \frac{ w_p \vec{u}_p\vec{u}_p }{c^2} + \delta_{ij} p_p \Big) \nonumber \\
&&+ \gamma_p n_p q_p (\vec{E}+\frac{\vec{v}_p}{c}\times\vec{B})
- \tau_{fr} n_p n_e (\vec{u}_p-\vec{u}_e),
\end{eqnarray}
\begin{eqnarray}
\label{eq:ene}
&& \pp{}{t} \Big(\gamma_p^2 w_p - p_p \Big)
= -\div ( \gamma_p w_p \vec{u}_p  ) \nonumber \\
&&+ \gamma_p n_p q_p (\vec{v_p}\cdot\vec{E})
- \tau_{fr} n_p n_e c^2 (\vec{\gamma}_p-\vec{\gamma}_e),
\end{eqnarray}

\begin{eqnarray}
\label{eq:B}
\pp{\vec{B}}{t} &=& - c \rot \vec{E} - \grad \psi , \\
\label{eq:E}
\pp{\vec{E}}{t} &=& c \rot \vec{B} - 4\pi \vec{j} - \grad \phi ,
\end{eqnarray}
\begin{eqnarray}
\label{eq:psi}
\pp{\psi}{t} &=& - c^2 (\div \vec{B}) - \kappa \psi , \\
\label{eq:phi}
\pp{\phi}{t} &=& - c^2 ( \div \vec{E} - {4\pi} \rho_c ) - \kappa \phi .
\end{eqnarray}
In these equations, $\gamma_s$ is the Lorentz factor,
$n_s$ is the proper density,
$\vec{u}_s=\gamma_s\vec{v}_s$ is the fluid 4-velocity,
$w_s=n_smc^2+[\Gamma/(\Gamma-1)]p_s$ is the enthalpy
with the specific heat ratio $\Gamma=4/3$,
$\delta_{ij}$ is the Kronecker delta,
$p_s$ is the proper pressure,
$q_p=-q_e$ is the positron/electron charge,
$\vec{j} = q_p n_p \vec{u}_p + q_e n_e \vec{u}_e$ is the electric current, and
$\rho_c=\gamma_p n_p q_p + \gamma_e  n_e q_e$ is the charge density.
In the momentum and the energy equations,
inter-species friction terms with the coefficient $\tau_{fr}$ are added.
They are revised from our previous work.
The new form is similar to a covariant form in \citet{guro86}.
The variables $\psi$ and $\phi$ are
virtual potentials for divergence cleaning \citep{munz00,dedner02,kom07},
whose propagation speed and the decay rate are $c$ and $\kappa$, respectively.

We study two-dimensional system evolutions in the $x$--$z$ plane.
We use the Harris-like model as an initial configuration:
the magnetic field, the (proper) density, and the pressure are
$\vec{B}(z) = B_{0} \tanh(z/L)~\vec{\hat{x}} + B_G~\vec{\hat{y}}$,
$n_{s}(z) = n_0 \cosh^{-2}(z/L) + n_{in}$, and
$p_{s}(z) = n_{s}(z)mc^2$, respectively.
Here, $L$ is the current sheet thickness,
$B_G$ is the guide field amplitude, and
$n_{in}=0.1 n_0$ is the background proper density.
The initial out-of-plane current is sustained by
the fluid drift of $\sim \pm 0.1c$.
This restricts
the electron inertial length (${\sim}0.1L$)
and the Larmor radius (${\lesssim}0.05L$) to sufficiently small values.
We set the reconnection point at the origin $(x,z) \sim (0,0)$
in the system domain of $[0, 120L] \times [-60L,60L]$
or $[0, 240L] \times [-60L,60L]$.
In order to reduce the computational cost,
we assume that
the system is point-symmetric around the reconnection point:
$f(x,z) = f(-x,-z)$ or $-f(-x,-z)$ for all times.
The boundary conditions at $x=0$ are set accordingly.
At the other three boundaries,
we basically consider
the Neumann boundary conditions ($\partial f / \partial \vec{n}=0$),
and the normal components of the fields ($B_n$ and $E_n$) are adjusted
to satisfy $\div \vec{B}=0$ and $\div \vec{E}=4\pi \rho_c$. 
We employed the spatially localized resistivity,
which is controlled by the coefficient $\tau_{fr}$. 
The effective Reynolds number of the frictional resistivity
is $S=30$ near the reconnection point
and $S=3000$ in the background region.
We normalize timescales by the light transit time $\tau_c=L/c$.
The decay time scale for divergence cleaning is set to $\kappa^{-1}=0.2 \tau_c$.

We studied the guide field effect
by changing $B_G$ as shown in Table \ref{table}.
We consider the magnetization parameters
$\sigma = B^2/[4\pi (2\gamma^2 w)]$,
the ratio of the Poynting flux to the plasma energy flux.
In the inflow region, $\sigma_{in} \approx \sigma_{x,in} + \sigma_{y,in}$,
where $\sigma_{x} = B_x^2/[4\pi (2\gamma^2 w)]$ and
$\sigma_{y} = B_y^2/[4\pi (2\gamma^2 w)]$ are
contributions from $B_x$ and $B_y$.
The parameter $\sigma_{x,in}$ is set to $4$ and
the relevant Alfv\'{e}n velocity is
$c_{A,in}=[\sigma_{x,in}/(1+\sigma_{x,in})]^{1/2}c=0.894c$.
The other parameter $\sigma_{y,in}$ depends on $B_G$.

The system evolution is solved by the modified Lax--Wendroff scheme
with a small artificial viscosity,
whose viscous coefficient depends on the local gradient of the fluid 4-velocity. 
We directly solve quartic equations to recover the primitive variables \citep{zeni09a}.
The results are checked by changing the system size,
the left boundary conditions (with or without the point symmetry),
the time and spatial resolutions,
and parameters for the numerical stability.

\begin{deluxetable}{l|ccccr}
\tabletypesize{\scriptsize}
\tablecaption{\label{table} List of Simulation Runs}
\tablewidth{0pt}
\tablehead{
\colhead{Run} &
\colhead{Domain Size} &
\colhead{Grid Points} &
\colhead{$B_G/B_{0}$} &
\colhead{$\sigma_{y,in}$} &
\colhead{$\mathcal{R}$}
}
\startdata
1 & 120 $\times$ 120 & 2400 $\times$ 2400 & 0.0 & 0.0 & 0.142 \\
2 & 120 $\times$ 120 & 2400 $\times$ 2400 & 0.25 & 0.25 & 0.126 \\
3a & 120 $\times$ 120 & 2400 $\times$ 2400 & 0.5 & 1.0 & 0.102 \\
3b & 240 $\times$ 120 & 4800 $\times$ 2400 & 0.5 & 1.0 & 0.102 \\
4 & 120 $\times$ 120 & 2400 $\times$ 2400 & 1.0 & 4.0 & 0.074 \\
5 & 120 $\times$ 120 & 2400 $\times$ 2400 & 1.5 & 9.0 & 0.055
\enddata
\tablecomments{
The initial guide field $B_G$,
the ratio of the its Poynting flux to the plasma energy flux $\sigma_{y,in}$,
and typical steady reconnection rates $\mathcal{R}$. \\
}
\end{deluxetable}

\section{SIMULATION RESULTS}

\begin{figure*}[htbp]
\begin{center}
\ifjournal
\includegraphics[width={\columnwidth},clip]{f1.eps}
\else
\includegraphics[width={2\columnwidth},clip]{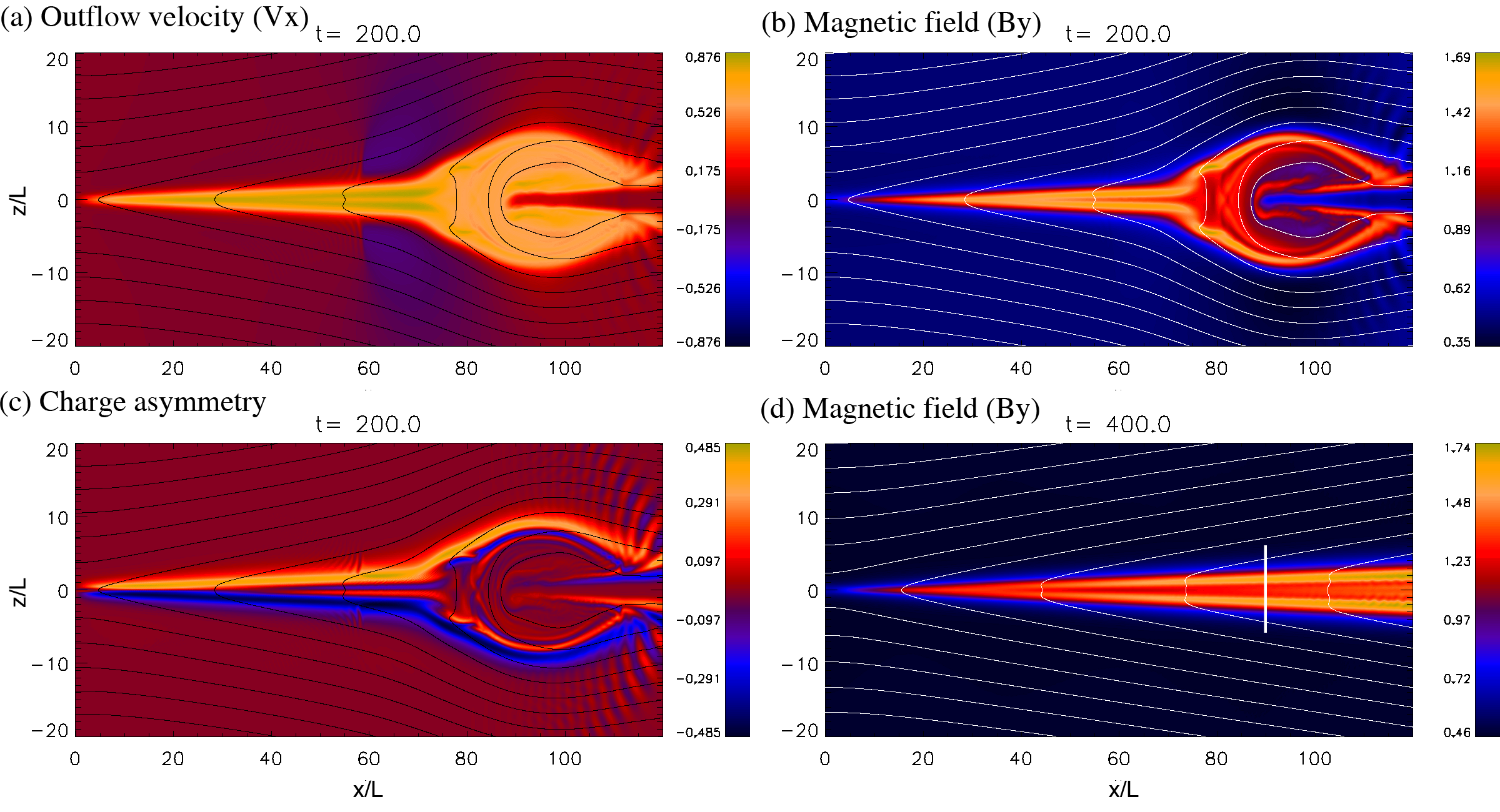}
\fi
\caption{
(Color online)
Snapshots of run 3b at $t=200\tau_c$ in the $x$--$z$ two-dimensional plane.
Contour lines show the in-plane magnetic fields.
(\textit{a}) The average plasma outflow
$\langle v_x\rangle = (n_pu_{x,p}+n_eu_{x,e})/(\gamma_pn_p+\gamma_en_e)$,
(\textit{b}) the out-of-plane magnetic field $B_y/B_0$, and
(\textit{c}) the charge separation $(\gamma_pn_p-\gamma_en_e)/(\gamma_pn_p+\gamma_en_e)$.
(\textit{d}) The out-of-plane magnetic field $B_y/B_0$ in run 3b at $t=400\tau_c$.
We discuss properties along the white line in Section \ref{section:outflow}.
\label{fig:snap}}
\end{center}
\end{figure*}

\subsection{Overview}

Snapshots of runs 3b with a guide field $B_G/B_0=0.5$
are presented in Figure \ref{fig:snap}.
Shown in Figure \ref{fig:snap}{\itshape a} is an outflow structure of
the plasma mean number flow,
\begin{eqnarray}
\langle\vec{v}\rangle=\frac{n_p\vec{u}_p+n_e\vec{u}_e}{\gamma_pn_p+\gamma_en_e}
.
\end{eqnarray}
We see that a fast outflow jet travels
to the magnetic island (plasmoid) in the downstream region.
Since there are ambient current sheet plasmas,
the plasmoid exhibits a crab claw structure in the downstream,
similar to large-scale MHD simulations
of nonrelativistic magnetic reconnection (e.g., \citet{shuei01}).
Inside the outflow channel,
the flow speed $\langle v_x \rangle$ is $0.8 \sim 0.85c$.
The maximum 4-velocity of each species is $u_{xs}\sim 2.1c$.
These speeds are slightly slower than the antiparallel counterpart.
The backward flows around $x \sim 70 L$ (blue regions; $\langle v_x\rangle<0$)
are reverse flow from the plasmoid.
Since the downstream plasmoid expands,
the surrounding magnetic fields are compressed
and then its magnetic pressure expels the plasma along the field lines.
Sharp boundaries at $x \sim 60 L$ are
the fronts of such backward flows.

Shown in Figure \ref{fig:snap}{\itshape b} is
the out-of-plane magnetic field $B_y/B_0$.
We see that the guide field is compressed
at the edge of the plasmoid
and in the narrow reconnection outflow region,
because the incoming reconnection flows
transport and pile up the $B_y$ magnetic flux from the upstream region.
Such $B_y$-compression changes
the composition of outgoing energy flux
in relativistic magnetic reconnection.

Figure \ref{fig:snap}{\itshape c} shows the charge separation,
$(\gamma_pn_p-\gamma_en_e)/(\gamma_pn_p+\gamma_en_e)$.
The charge density $\rho_c$ looks similar.
We see that positrons dominate
in the upper of the outflow region,
and electrons in the lower.
The charge separation generates
the electric fields in the $x$--$z$ plane:
$E_x<0$ in the upper half of the inflow region,
$E_x>0$ in the lower half of the inflow region, and
$E_z<0$ in the rightward outflow region.
The $\vec{E}\times\vec{B}$ condition with the guide field $B_y$
is consistent with the reconnection flow pattern.
It is impressive to see that
the non-neutral layers extend over the large spatial domain
and that the charge separation is strong $\sim$0.5.
Such a non-neutral structure is qualitatively consistent with
positron-rich or electron-rich regions
around the reconnecting diffusion region in previous kinetic work
\citep{zeni08}.\footnote{Charge signs are opposite from
Figure 2(c) in \citet{zeni08}, because the guide field is set oppositely.}
Stripes around $x \lesssim 60L$ are transient oscillations invoked by the reverse flow.
As the system evolves, the plasmoid and the reverse flows move further away.
We also see strong perturbations around $x \sim 120L$,
where the plasmoid hits the ambient plasmas.

Later, the system approaches to a quasi-steady stage.
Shown in Figure \ref{fig:snap}{\itshape d} is
a late-time structure of $B_y$ at $t=400\tau_c$ in run 3b.
We see that the outflow structure
in Figure \ref{fig:snap}{\itshape b}
straightforwardly extends toward the $+x$ direction.
The backward flow fronts are outside the presented domain at this time.
The in-plane magnetic fields show a narrow Petschek-type structure.
Interestingly, the $B_y$-structure is weakly bifurcated, and then
the two peaks separate the outflow channel into three layers.
We examine the outflow channel structure in Section \ref{section:outflow}.
Around $x\sim 115L$, the bifurcated peaks
exhibit a very weak wave structure,
probably due to a velocity--shear driven instability.
This introduces a minor oscillation inside the central channel
in the further downstream; however,
it does not change the global outflow structure.
We also confirmed that
the global outflow structure does not depend on the domain size
by comparing runs 3a and 3b.
Therefore, the outflow boundary condition does not change the physics.

We believe that the large distance to the upstream boundary ($z=60L$) is
virtually eliminates boundary effects on the simulation.
The plasmoid and reconnected field lines are well confined
in $|z| < 10$--$20L$ in panels in Figure \ref{fig:snap}.
In the steady stage of $t=400\tau_c$,
the slope angle of the field lines is only $\sim 0.1$ around the outflow region,
and it becomes even smaller further upstream.
For example, in the vicinity of the top boundary,
the field line at $(0,+56.2L)$ is connected to $(120L,+60L)$.

\subsection{Reconnection electric field}

\begin{figure*}[thbp]
\begin{center}
\ifjournal
\includegraphics[width={\columnwidth},clip]{f2.eps}
\else
\includegraphics[width={2\columnwidth},clip]{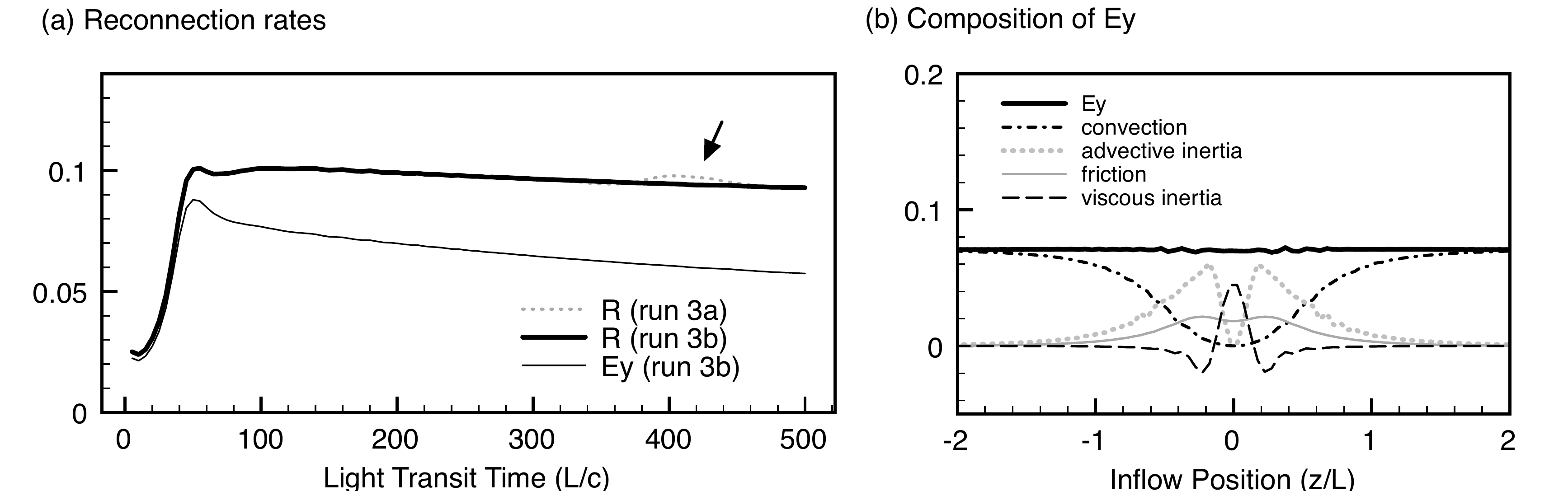}
\fi
\caption{
(\textit{a})
Time evolution of reconnection rates:
normalized rates $\mathcal{R}$ for runs 3a and 3b,
and a raw reconnection rate for run 3b $E_y/B_0$.
(\textit{b})
Composition of the reconnection electric field $E_y$
along the inflow line ($x=0$) in run 3b at $t=200\tau_c$:
the convection term, the advective inertia, the frictional resistivity,
and the viscous inertia,
as presented in Equation \ref{eq:ohm_y}.
These values are normalized by $B_0$.
\label{fig:Ey}}
\end{center}
\end{figure*}

The reconnection electric field $E_y$ at the $X$-point
presents the transfer rate of
the antiparallel magnetic fields from the upstream region.
Since the surrounding conditions change in time,
the normalized rate or ``reconnection rate''
gives a good measure of the system evolution.
Here we define the rate as
\begin{eqnarray}
\mathcal{R}=\frac{c E_{y} }{ c_{A,in'} |B_{x,in'}| },
\end{eqnarray}
where the subscript $in'$ denotes the upstream properties
at $x=0, z=20L$.
The time evolution of the rates is shown in Figure \ref{fig:Ey}{\itshape a},
and we see that the (normalized) reconnection becomes constant
after the initial ramp up for $t \gtrsim 50\tau_c$. 
For comparison, the rate for run 3a is presented, too.
Runs 3a and 3b are in excellent agreement
except for around $t\sim 400\tau_c$ (indicated by an arrow)
when the boundary effect temporally comes back to the reconnection point
in the smaller run 3a.
In Table \ref{table}, we present
the peak values of the quasi-steady reconnection rates
for all other runs.

So, what sustains such quasi-steady evolution of magnetic reconnection?
In other words, what is responsible for
the reconnection electric field $E_y$
around the $X$-point in our simulations?
We study the composition of $E_y$ near the $X$-point using the Ohm's law.
Combining the momentum equations (Equation \ref{eq:mom}) of the two species,
we obtain the following relation:
\begin{eqnarray}
\label{eq:ohm}
\vec{E}
&=&
- \frac{\langle\vec{v}\rangle}{c}\times\vec{B}
+ \frac{1}{(\gamma_p n_p+\gamma_e n_e)q_p}
\nonumber \\
&&
\Big[
m\gamma_p n_p
( \vec{v}_p \cdot \grad ) h_p\vec{u}_p
-
m \gamma_e n_e
( \vec{v}_e \cdot \grad ) h_e\vec{u}_e
\nonumber \\
&&
~~~ + (\grad p_p - \grad p_e)
+ {2\tau_{fr}n_pn_e}(\vec{u}_{p}-\vec{u}_{e})
\Big]
,
\end{eqnarray}
where $h_s=w_s/(n_smc^2)$ is the specific enthalpy.
Note that we drop the time derivatives
considering the quasi-steady condition ($\partial_t\sim 0$).
In addition, along the inflow line ($x=0$),
we find $n_p = n_e$, $\gamma_p = \gamma_e$,
$v_{z,p} = v_{z,e} = \langle{v_z}\rangle$,
$h_{p}u_{y,p} = - h_{e}u_{y,e}$,
and $B_z = 0$, and so 
the equation can further be simplified.
We can also approximate that
the artificial viscosity works for $h_s\vec{u}_{s}$
in a form of $-\nu_z {\partial_{zz}}$
with an effective viscous coefficient $\nu_z$. 
The $x$-derivative terms (${\partial}_x$ and ${\partial_{xx}}$)
are negligible in this case. 
As a result, the $y$-component of Equation \ref{eq:ohm} along the inflow line yields
\begin{eqnarray}
\label{eq:ohm_y}
E_y
&\approx&
\frac{-\langle v_{z}\rangle B_x}{c}
+ \frac{m\langle{v_z}\rangle}{q_p}
\frac{\partial h_pu_{y,p}}{\partial z}
+ \frac{\eta_{eff}}{\gamma_p} {j_y}
- \frac{m\nu_z}{q_p}
\frac{\partial^2 h_pu_{y,p}}{\partial z^2}
,
\nonumber \\
\end{eqnarray}
where $\eta_{eff}=(\tau_{fr}/q^2_p)$ is
the effective resistivity by the inter-species friction.
Terms in the right-hand side represent
the field convection, the advective inertia of the fluid,
the frictional resistivity, and the viscous inertia, respectively.

Based on Equation \ref{eq:ohm_y},
we study the composition of the reconnection electric field $E_y$
in Figure \ref{fig:Ey}{\itshape b}.
Here, $\nu_z$ is represented by the typical value around the neutral point. 
Outside the central diffusion region,
the convection term mostly explains the electric field.
The frictional resistivity term works
inside the diffusion region where the current is strong.
However, it explains only a quarter of $E_y$.
We find that
the local Lorentz factor $\gamma_p{\sim}1.6$ partially suppresses
this term in Equation \ref{eq:ohm_y}.
Instead, the other two inertial effects are important.
The advective term is the biggest contributor.
In addition to the acceleration in the ${\pm}y$ directions ($u_{y,s}$),
the specific enthalpy $h_s\sim 1+4p_s/(n_smc^2)$ increases
because dissipation process heats incoming plasmas.
In a very vicinity of the neutral plane ($z=0$),
the viscous inertial effect replaces the advection.
Although we introduced it for the numerical stability,
we believe it reasonably represent the physics
because the kinetic effect plays a quasi-viscous role
near the neutral plane \citep{hesse04}.

We carried out a similar analysis
in the antiparallel case (run 1), too.
The frictional resistivity plays a bigger role,
but it explains only one-third of the electric field.
The local Lorentz factor $\gamma_p{\sim}1.6$ similarly
suppresses the frictional resistivity.
Regardless of the presence of the guide field,
the inertial effects play a role to
sustain magnetic reconnection in the relativistic two-fluid system.

\subsection{Energy budget}

Next, we investigate
how the guide field changes
the energy conversion process near the reconnection region
in the quasi-steady stage.
We consider the composition of
the energy flux $\vec{F}$ in the following way:
\begin{eqnarray}
\label{eq:flux}
\vec{F}
&=&
\frac{c\vec{E}\times\vec{B}}{4\pi}+\sum_{s=p,e}\gamma_sw_s\vec{u}_s \\
\label{eq:flux2}
&=&
\frac{c\vec{E}\times B_y\vec{\hat{y}}}{4\pi}
+ \frac{c \vec{E}\times
( B_x\vec{\hat{x}} + B_z\vec{\hat{z}} ) }{4\pi}
+ \sum_s \frac{\Gamma \gamma_sp_s }{\Gamma-1} \vec{u}_{s}
\nonumber \\
&&
+ mc^2 \sum_s (\gamma_s-1)n_s \vec{u}_{s}
+ mc^2 \sum_s n_s \vec{u}_{s}
.
\end{eqnarray}
In Equation \ref{eq:flux2},
we decompose
the Poynting flux into
the contribution by the guide field ($B_y$) and
the one by the reconnecting magnetic fields ($B_x$ and $B_z$).
The plasma energy flux is decomposed into three parts.
The pressure term transports the gas internal energy and
it also contains the work by the gas.
Since it recovers the classical enthalpy flux
in the nonrelativistic limit
($\rightarrow \sum_s \frac{5}{2} p_s\vec{v}_{s}$ with $\Gamma=5/3$),
we call this term ``enthalpy flux'' in this work.
Note that the rest-mass contributions are considered separately
from this enthalpy.
The last two stand for
the bulk kinetic energy
($\rightarrow \sum_s \frac{1}{2} mn_s{v}_s^2\vec{v}_{s}$)
and the matter flow
($\rightarrow mc^2 \sum_s n_s\vec{v}_{s}$), respectively.

\begin{figure}[htbp]
\begin{center}
\ifjournal
\includegraphics[width={\columnwidth},clip]{f3.eps}
\else
\includegraphics[width={\columnwidth},clip]{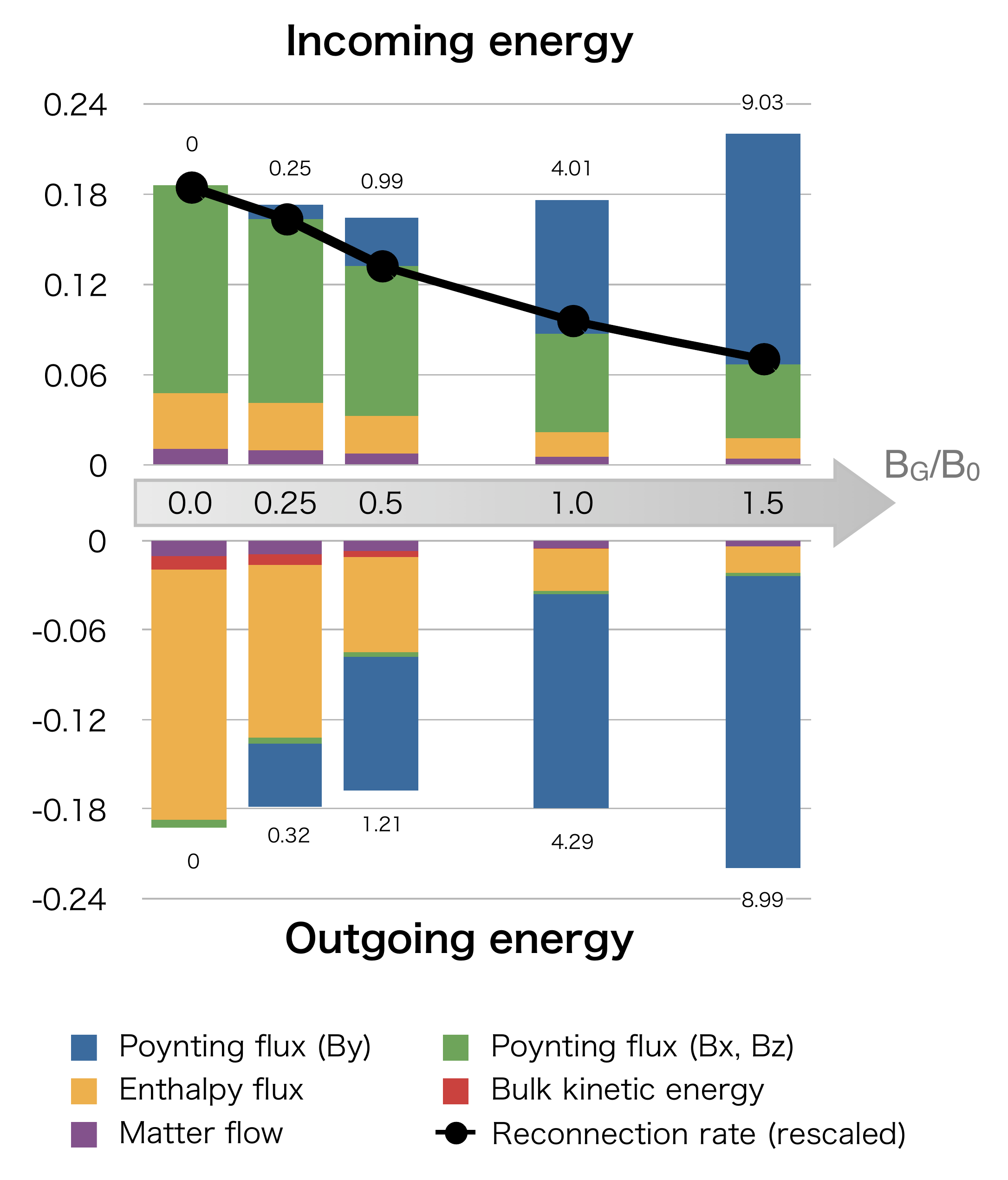}
\fi
\caption{
(Color online)
The incoming and outgoing energy fluxes
around the reconnection region ($|z|<10L,|x|<40L$).
The guide field Poynting flux,
the rest part of Poynting flux,
the plasma enthalpy flux,
the bulk kinetic energy,
and the matter flow are presented,
as presented in Equation \ref{eq:flux2}.
The reconnection rates $\mathcal{R}$ are rescaled from Table \ref{table}.
The relevant $\sigma_y$ parameter is indicated by small numbers
on the top/bottom of the bars.
\label{fig:energy}}
\end{center}
\end{figure}

We consider the energy budget
around the reconnecting square region of $|z|<10L,|x|<40L$.
The top panel in Figure \ref{fig:energy} presents
the composition of the incoming energy flux
($-\vec{F}\cdot \vec{\hat{z}}$)
per cross section at the inflow boundary ($z=10L$)
as a function of the guide field.
They are normalized by the typical Poynting flux $c(B^2_{x}/4\pi)$
at the upstream border $x=0,z=10L$.
The bottom panel in Figure \ref{fig:energy} presents
the outgoing energy flux ($-\vec{F}\cdot \vec{\hat{x}}$)
at the outflow boundary ($x=40L$),
evaluated by the same unit as the upstream values.
They are evaluated at the specific time steps in different runs,
but they are carefully selected
so that the energy budget becomes nearly steady in the rectangle region.
We see that both two fluxes are well balanced.
Although it may be difficult to recognize in the figure,
the matter fluxes are also well balanced.

For comparison,
rescaled values of reconnection rates (Table \ref{table})
are overplotted on the top panel in Figure \ref{fig:energy}.
It is reasonable to see that
they are proportional to the incoming energy flux
except for $B_y$-Poynting flux.
The reconnection rate is relevant to
the inflow speed $v_{in}$, which transports
the upstream antiparallel magnetic flux $B_x$
into the reconnected field $B_z$.
In these cases, under the similar upstream conditions,
the $B_x$-Poyting flux decreases as ${\sim}\mathcal{R}$ and
the accompanying plasma flux also decreases accordingly.
We also see that the reconnection rate decreases but
it weakly depends on the guide field amplitude $B_G/B_0$.
This trend is consistent with
a lot of previous nonrelativistic surveys (e.g., \citet{huba05}).

We see that reconnection well dissipates
the upstream antiparallel magnetic energy.
When the guide field is weak,
the Poynting energy is mostly converted into the plasma energy.
Importantly,
the energy is converted to the enthalpy flux
rather than the bulk kinetic energy.
The plasma pressure becomes strong in the outflow region
in order to balance the strong upstream magnetic pressure.
In magnetically dominated cases,
plasma temperature becomes relativistic,
and then such relativistic pressure substantially enhances the enthalpy flow.
Therefore, the enthalpy flux dominates the bulk kinetic energy.

As the guide field increases,
the relevant Poynting flux becomes a major component of the incoming energy flux.
Simultaneously, the outgoing energy flux is dominated by
the Poynting flux of the guide field.
In the case of $B_G/B_0=0.5$ (run 3),
as presented in Section 3.1,
the guide field is compressed by a factor of ${\sim}3$.
Consequently, the outgoing guide field Poynting flux increases and then
it exceeds the enthalpy flux.
When the guide field is stronger,
the outflow is dominated by the Poynting flux by $B_y$.
In the outflow region,
the ratio of the $B_y$-Poynting flux to
the plasma energy flux is $\sigma_{y,out} = 0, 0.32, 1.21, 4.29$, and $8.99$,
respectively.
It is interesting to see that
these ratios are similar to
the initial upstream $\sigma_{y,in}$ parameter (Table \ref{table}).

Under the single-fluid ideal RMHD approximation,
we have the following relations
from the polytropic law, the continuity, and
the out-of-plane flux conservation:
\begin{eqnarray}
\frac{d}{dt}\Big(\frac{p'}{n'^\Gamma}\Big) = 0, ~~
\frac{d}{dt}\Big(\frac{B_y}{\gamma' n'}\Big)=0,
\end{eqnarray}
where the prime sign ($'$) denotes the single-fluid properties.
They immediately suggest
that $\sigma_y$ increases
through the weak compression by magnetic reconnection,
\begin{eqnarray}
\frac{d \ln \sigma_y}{d t}
\sim \frac{d}{dt} \ln \frac{ (B_y/\gamma')^2 }{16\pi p' }
= (2-\Gamma) \frac{d \ln n'}{dt}
.
\end{eqnarray}
In the case of run 3b, the typical compressional ratio of $\sim 2$
suggests that $\sigma_y$ increases by a factor of $\sim 1.6$.
However, since our two-fluid system contains non-ideal effects,
$\frac{d}{dt} (p'/n'^{\Gamma}) > 0$ tends to reduce $\sigma_{y}$.
In the stronger guide field cases,
it is known that plasmas behave incompressibly and so
the system is more likely to preserve $\sigma_{y,in} \sim \sigma_{y,out}$.

\subsection{Outflow channel}
\label{section:outflow}

\begin{figure}[htbp]
\begin{center}
\ifjournal
\includegraphics[width={0.6\columnwidth},clip]{f4.eps}
\else
\includegraphics[width={\columnwidth},clip]{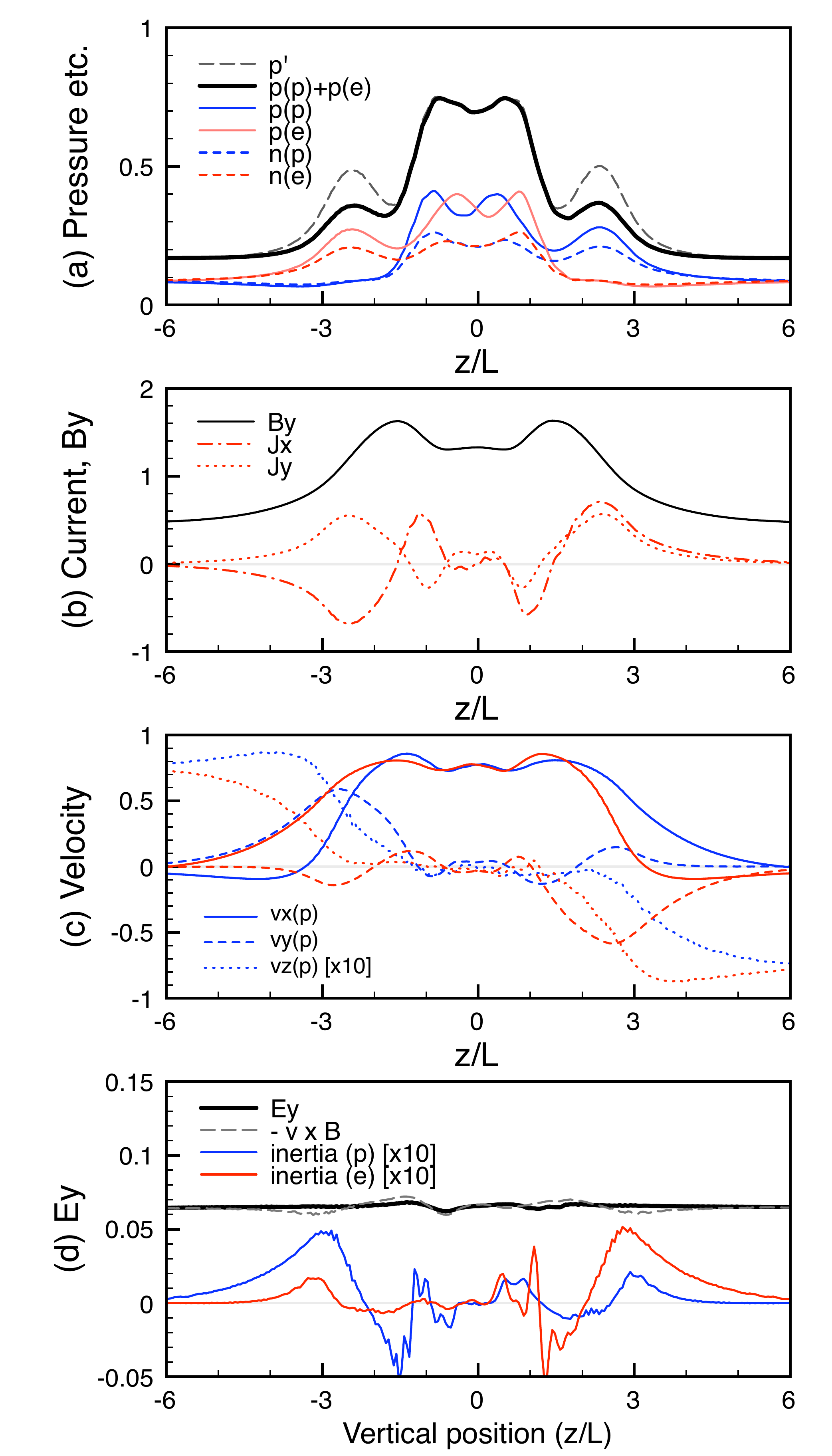}
\fi
\caption{
(Color online)
Structure of the outflow channel
at $x=90L$ at $t=400\tau_c$.
(\textit{a}) Proper pressure $p_p$, $p_e$,
positron density $n_p$, $n_e$,
total plasma pressure $p_p+p_e$, and
the relevant single-fluid pressure $p'$ are presented.
(\textit{b}) The out-of-plane magnetic field $B_y$, and
the electric currents $j_x$ and $j_y$.
(\textit{c}) Fluid velocities, $v_{x,p}, v_{x,e}, v_{y,p}, v_{y,e}, v_{z,p}$, and $v_{z,e}$.
The $z$-components are magnified 10 times larger.
(\textit{d}) Composition of the electric field, $E_y$,
$(-\langle\vec{v}\rangle \times \vec{B}/c)_y$, and
the $z$-convection of the inertial terms
(Equations \ref{eq:vz_inertia_p} and \ref{eq:vz_inertia_e}).
The inertial terms are magnified 10 times larger.
\label{fig:1Dcut}}
\end{center}
\end{figure}

We look at the structure of the outflow region.
Shown in Figure \ref{fig:1Dcut} are
one-dimensional profiles across the outflow channel,
at $x=90L$ at $t=400\tau_c$.
The cut line is indicated by
a white line in Figure \ref{fig:snap}{\itshape d}.
From these profiles,
we see that the outflow region consists of three characteristic layers,
separated by two peaks of $B_y$:
(1) a positron-rich boundary layer on the upper side ($2L \lesssim z \lesssim 4L$),
(2) a central channel with high plasma pressure, and
(3) an electron-rich boundary layer on the lower side.

The central channel contains a minor oscillating structure
in plasma pressure (Figure \ref{fig:1Dcut}{\itshape a}).
Inside the central channel, the plasma temperature is hot
and the outgoing fluid velocities are roughly similar
$v_{s,x}\sim 0.8c$ across the channel.
Since the Lorentz force $q_s(v_{s,x}B_y)$ works differently
in the presence of the guide field $B_y$,
the positron and electron outflow channels are oppositely displaced to the ${\pm}z$-directions.
The structures of the density, the pressure,
the outflow speed $v_x$, and the inflow speed $v_z$
(magnified 10 times larger in Figure \ref{fig:1Dcut}{\itshape c})
are all consistent with the displacement of the flow channels.
The central channel becomes wider in the further downstream and
we confirm that the system keeps the three-layer structure at least $x\sim 180L$.

We see that currents in the boundary layers are responsible
for the global magnetic field topology.
For example, $j_x$ explains the compression of the guide field $B_y$, and
a ``bifurcated'' $j_y$-structure is consistent with
the Petschek-type structure of in-plane magnetic fields
(Figure \ref{fig:snap}{\itshape d}).
We also see minor reverse currents near the center, but
they are less important.
Note that relative motion between the two species is significant
in the boundary layers (Figure \ref{fig:1Dcut}{\itshape c}),
because plasmas sustain the currents in low-dense regions.
Interestingly, in the upper {\itshape positron}-rich boundary layer,
we see a fast {\itshape electron} flow in the $y$-direction
and indeed electrons are major contributors to $j_y$ there.
This is because the Lorentz force $j_x B_z$ modulates
the local bulk outflow to the $-y$-direction, and so
the electron flow looks overemphasized in our simulation frame.

In addition, in order to see the difference between
our two-fluid model and the single-fluid RMHD model,
we evaluate an equivalent single-fluid properties in the following way.
We translate the conserved properties
with the single-fluid properties (with the prime sign),
\begin{eqnarray}
\label{eq:2den}
\gamma_p n_p +\gamma_e n_e &=& \gamma' n' , \\
\label{eq:2mom}
\gamma_pw_p\vec{u}_p+\gamma_ew_e\vec{u}_e &=& \gamma' w' \vec{u}' , \\
\label{eq:2ene}
\gamma_p^2w_p - p_p +\gamma^2_ew_e - p_e &=& \gamma'^2 w' - p' ,
\end{eqnarray}
and then we calculate the single-fluid primitive variables
by solving a quartic equation \citep{zeni09a}.
The single-fluid pressure is assumed to be isotropic.
Combining two fluid pressure into a single-fluid pressure
does not lead to an isotropic pressure unless the relative velocity vanishes.
The obtained pressure $p'$ is
overplotted in Figure \ref{fig:1Dcut}{\itshape a}.
We see that $p'$ is in excellent agreement with
the total two-fluid pressure $(p_p+p_e)$ in the central channel.
On the other hand,
the discrepancy is significant $p'/(p_p+p_e){\sim}1.4$ in the two boundary layers
and so it may widen the outflow channel in single-fluid simulations.
It is reasonable that the discrepancy is observed
where the relative motion between the species is significant
(Figure \ref{fig:1Dcut}{\itshape c}).

In the boundary layers the local frozen-in condition also breaks down.
We similarly analyze the composition of the reconnection electric field $E_y$
based on Equation \ref{eq:ohm}.
Figure \ref{fig:1Dcut}{\itshape d} presents the profile of $E_y$,
along with the field convection and
the $z$-convection of fluid inertia terms,
\begin{eqnarray}
\label{eq:vz_inertia_p}
\frac{m\gamma_p n_p}{q_p(\gamma_p n_p+\gamma_e n_e)}
~{v}_{z,p}\pp{h_pu_{y,p}}{z},\\
\label{eq:vz_inertia_e}
-\frac{m\gamma_e n_e}{q_p(\gamma_p n_p+\gamma_e n_e)}
~{v}_{z,e}\pp{h_eu_{y,e}}{z}.
\end{eqnarray}
In Figure \ref{fig:1Dcut}{\itshape d},
the inertia terms are magnified
10 times larger to emphasize them.
We found that the non-ideal contribution
(the difference between $E_y$ and
the $\langle\vec{v}\rangle\times\vec{B}$ term)
is mostly explained by these terms.
The positron inertial term appears in the lower electron-rich boundary layer,
and electron term in the lower electron-rich layer.
For example, in the positron-rich boundary region,
both the $z$-component of the positron velocity and the $y$-momentum of the positrons
are much smaller than that of the electrons.
As a result, the electron inertial term dominates
even though the positron density is larger.
Around $z\sim\pm 3 L$,
the inertial contributions from both two fluids sustain 12\% of $E_y$. 
Because of the large Reynolds number $S \sim 3000$,
the frictional resistivity effect is negligible and
we confirmed that the viscosity is negligible, too. 
Compared with the reconnection site
where the nonideal terms are 100\% responsible for the electric field,
the nonideal effects are smaller, but they still play a role here.

\begin{figure}[htbp]
\begin{center}
\ifjournal
\includegraphics[width={0.6\columnwidth},clip]{f5.eps}
\else
\includegraphics[width={\columnwidth},clip]{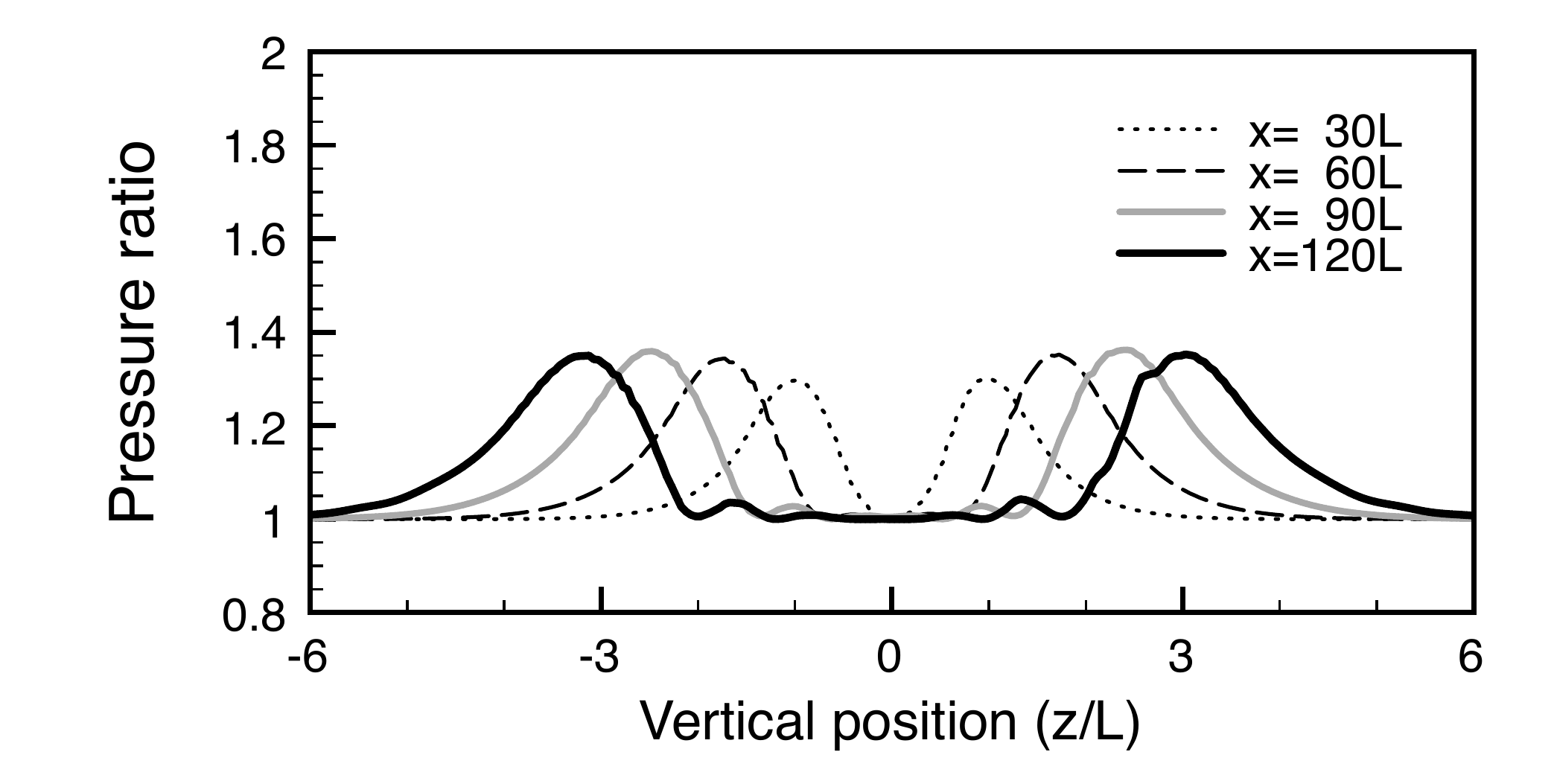}
\fi
\caption{
One-dimensional profiles of $p'/(p_p+p_e)$,
the ratio of the single-fluid pressure
to the total two-fluid pressure.
They are taken from $x=30,60,90,$ and $120L$ at $t=400\tau_c$ in run 3b.
\label{fig:xp}}
\end{center}
\end{figure}

Shown in Figure \ref{fig:xp} is
the $x$-dependence of the ratio of the single-fluid pressure ($p'$)
to the total fluid pressure ($p_p+p_e$),
which is a good measure of the two-fluid effects.
As the central outflow channel expands downstream,
the two-fluid effects are always localized in the relevant boundary regions.
Their amplitude and spatial width are similar
until the backward flows around the plasmoid hit the boundary layers.

\section{DISCUSSION AND SUMMARY}

Regarding basic models of antiparallel reconnections,
there have been two controversial opinions on the relativistic reconnection rate.
\citet{bf94b} argued that relativistic reconnection can involve
relativistic inflow and the fast consumption of the upstream magnetic energy,
because Lorentz contraction of the plasma outflow enables
larger energy output per cross section.
On the other hand, \citet{lyu05} pointed out
that relativistic pressure increases an effective inertia $w$,
and that the shock balance condition allows a narrower Petschek outflow.
Therefore, he argued that
relativistic reconnection will not be
an efficient energy converter,
due to the slow outflow and the narrower energy output channel.
Numerically, \citet{zeni09a} demonstrated that
reconnection tends to be faster and faster in magnetically dominated regimes,
although the obtained Petschek-type structure is
best described by the \citet{lyu05} model.
The unsolved problem is
what makes reconnection faster, or
what balances faster energy input?

In our energy balance analysis,
we found that
the outgoing plasma energy is mostly carried
by
the enthalpy flux or the internal pressure flux
in the antiparallel case. 
We think this is an important reason
why relativistic magnetic reconnection is faster.
The plasma pressure in the outflow region needs to be relativistic
in order to balance a strong upstream magnetic pressure \citep{lyu05}.
As the outflow plasma temperature becomes relativistically hot,
we see that the enthalpy flux (pressure part) vastly exceeds the other parts
by a factor of ${\sim}4( {p}/{nmc^2})$ in the outflow channel. 
Since the enthalpy flux carries larger energy per cross section,
the system can sustain faster energy input from the inflow region.
In other words, the enthalpy flux is a key to sustaining fast reconnection. 

The guide field introduced interesting changes to the reconnection structure.
Since no process can annihilate the out-of-plane magnetic field,
the inflows transport the upstream guide field to the narrow outflow region,
and therefore the guide field is compressed inside the outflow channel.
This significantly changes the energy composition in the reconnection outflow:
the outgoing energy flow is rather dominated by the Poynting flux
of the compressed guide field.
Although there is no theoretical proof,
our simulation results conserved (or slightly increased)
the ratio of the $B_y$-Poynting flux to the plasma energy $\sigma_y$.
If we employ a principle of $\sigma_y$ conservation,
we can crudely estimate that
the ratio of the upstream $B_x$-Poynting flux to
the downstream $B_y$-Poynting flux and the downstream enthalpy flow
is given by $1:\sigma_{y,in}$.

The boundary layers around the central reconnection jet
exhibit a significant charge separation
$(\gamma_pn_p-\gamma_en_e)/(\gamma_pn_p+\gamma_en_e) \sim 0.5$
to retain the electric field system.
The relative motion of species is significant in these layers, too.
We think these conditions are
beyond the scope of the single-fluid RMHD approach.
The present resistive RMHD codes 
solve the charge distribution $\rho_c$ and the current $\vec{j}$,
independently from plasma bulk motion.
Such an approximation is valid
when the charge separation and the relative motion are small.
In fact, in the boundary layers, we pointed out that
the single-fluid RMHD pressure and the two-fluid pressure differ
by a factor of $\sim 1.4$.
We also note that 
\citet{koide09} (see Section 7.1 for a summary) discussed
the validity of generalized RMHD equations in detail.
In our case of a pair plasma with relativistic pressure,
the pressure condition ($p_p/\rho_p \approx p_e/\rho_e$) is weakly violated
and the ``proper charge neutrality'' ($n_p \approx n_e$) breaks down
in the boundary layers (Figure \ref{fig:1Dcut}{\itshape a}).

We demonstrated that
the fluid inertial effects are important to
sustain magnetic reconnection in the reconnection region.
This is a characteristic feature of the two-fluid model.
By definition, a single-fluid RMHD model
only takes care of the fluid bulk inertia,
while it does not consider the inertial resistivity.
The inertia effects violate 
the ideal MHD condition around the bifurcated boundary layers, too.
In other words, inertial effects enhance an effective resistivity
both in the reconnection region and near the discontinuities,
playing a similar role as an anomalous-type resistivity.
To mimic such physics,
the nonrelativistic MHD simulations often employ
resistivity profiles based on the electric current,
for example, $\eta = \eta_0 + \eta_1 [\max(0,\vec{j}/\rho-v_{crit})]^2$,
but its relativistic extension may not be straightforward.

Using a single-fluid RMHD theory,
\citet{lyu05} predicted that
relativistic guide field reconnection
involves a three-layer structure in the outflow region,
with rotational discontinuities (outside) and slow shocks (inside).
Furthermore, his balance analysis yielded that
the outflow flux is dominated by the compressed guide field flux
between the rotational discontinuities.
In this work, we showed
a different three-layer structure in the outflow region:
the central hot outflow inside two charge-separated boundary layers. 
In our case,
the rotational discontinuities cannot appear
in the magnetically dominated upstream regions,
because the system runs out of plasma to sustain the electric current.
The upstream region does not have sufficient plasmas to sustain
the charge separation for the vertical electric field $E_z$, and therefore
the $B_y$-Poynting flux ($-cE_zB_y/4\pi$) is confined
around the central dense channel.
On the other hand,
the other important prediction qualitatively holds true 
---the outward energy is dominated by the guide field Poynting flux.
Energy transfer from the Poynting energy to the plasma energy
will take place somewhere in the downstream edge,
such as the interaction between plasmoids and the ambient medium.



In summary,
we carried out full two-fluid simulations of
relativistic magnetic reconnection with a guide field
and investigated the characteristic properties,
such as the charge separation and the guide field compression.
We demonstrated that
the guide field drastically changes
the composition of the output energy flux,
from enthalpy-dominated flow to Poynting-dominated flow,
potentially controlled by $\sigma_{y,in}$.
Inertial effects play a role of the effective resistivity,
and so they violate an ideal frozen-in condition. 
Most importantly, we showed that
the multi-fluid approach is very useful to
study important relativistic plasma problems.
It will be important to further study the consistency between
the single-fluid RMHD model and the multi-fluid model.

\begin{acknowledgments}
The authors are grateful to
R. Yoshitake and M. Kuznetsova for helpful comments. 
The authors also thank the anonymous referee for
his/her constructive comments on this manuscript.
This research was supported by the NASA Center for Computational Sciences,
and NASA's \textit{MMS} SMART mission.
S.Z. gratefully acknowledges
support from NASA's postdoctoral program.
\end{acknowledgments}


\end{document}